# Synchrotron Science for Sustainability: Life Cycle of Metals in the Environment

Louisa Smieska[1] [*]
Mary Lou Guerinot[2]
Karin Olson Hoal[3]
Matthew Reid[4]
Olena Vatamaniuk[5]

1. Cornell High Energy Synchrotron Source (CHESS), Cornell University, Ithaca NY 14853
2. Department of Biological Sciences, Dartmouth College, Hanover, NH, 03755
3. Department of Earth & Atmospheric Sciences, Cornell University, Ithaca NY 14853
4. Department of Civil and Environmental Engineering, Cornell University, Ithaca NY 14853
5. School of Integrative Plant Science Plant Biology Section, Cornell University, Ithaca NY 14853

* Corresponding author: 161 Synchrotron Drive, Ithaca, NY 14853, lmb327@cornell.edu

**Abstract**
The movement of metals through the environment links together a wide range of scientific fields: from earth sciences and geology as weathering releases minerals; to environmental sciences as metals are mobilized and transformed, cycling through soil and water; to biology as living things take up metals from their surroundings. Studies of these fundamental processes all require quantitative analysis of metal concentrations, locations, and chemical states. Synchrotron x-ray tools can address these requirements with high sensitivity, high spatial resolution, and minimal sample preparation. This perspective describes the state of fundamental scientific questions in the lifecycle of metals, from rocks to ecosystems, from soils to plants, and from environment to animals. Key x-ray capabilities and facility infrastructure for future synchrotron-based analytical resources serving these areas are summarized, and potential opportunities for future experiments are explored.

## I. Introduction

Overview:
Many critical converging questions surround the fundamentally intertwined mechanisms of elemental movement from geological contexts through soils, waterways, plants, and animals. Element transport and redox cycling on broad geological scales of space and time ultimately determine the local chemical resources for all living things, including the bioavailability of metals and micronutrients. These inorganic and organic mechanisms are essential to understanding key processes of life: how do living things regulate the uptake, transport, and storage of micronutrients they need, from whole-organism to single-cell scales, and where do these critical materials come from? How are metals and micronutrients transported to and used in cell processes, and how does the genetic makeup of life translate to its ability to thrive? Answers to these fundamental and linked questions underpin advances in the sustainability of mineral resourcing, agriculture, and managing the carbon and nitrogen cycles. Synchrotron-based techniques can address each of these areas.

Urgency:
Critical challenges to future global sustainability are many. They include the environmental impacts of human activities; effects of climate change on geochemical processes in soils and water; elemental transport in soils and living organisms, including uptake by plants; risks of materials scarcity for the renewable energy transition; and biosystems contamination and toxicity, ultimately compromising the already daunting task of feeding an estimated 11 billion people by 2100 [11]. These are fundamental science challenges that can be addressed by applying interdisciplinary tools, including high-precision analytical research, from the submicron to the industrial scale. A multidisciplinary, multi-investigator team-based approach is critical to addressing these important issues, across disciplines, functions, and biases. Large-scale infrastructure such as synchrotron light sources are inherently cross-disciplinary and collaborative environments, cultivating research, education, and innovation to address complex challenges [13].

Imminent developments:

These challenges are tied together not only by the need to understand the movement and chemical transformations of elements but also by novel experimental possibilities. The common thread of taking inventory of metals, their mineral and biogenic deportment, and their oxidation states across space and time runs through the approach laid out in this perspective [14]. Synchrotron light sources have long played a crucial role in advancing these fields by providing sensitive, quantitative, highly spatially resolved tools for measuring chemical composition of complex samples. In particular, x-ray fluorescence microscopy (XRF) provides ppm sensitivities for elemental imaging at spatial resolutions reaching below 1 µm (10s of nm), while x-ray spectroscopy probes the oxidation state and chemical bonding environment of a specific element [15]. As x-ray beams grow ever smaller and brighter, and instrumentation becomes faster and more flexible, there are both new questions we can ask and strive to answer, and new applications development for light sources that will facilitate convergent and interdisciplinary research on the life cycle of metals.

In this contribution, we identify four areas in which the future of synchrotron science will link data and knowledge across fundamental scientific questions in the lifecycle of metals, from rocks to ecosystems, animals and humans. These linkages are 1) metal transport through geological processes from the subsurface to soils, 2) elemental cycling mechanisms at the interface between soils and biogenic activity, 3) metals in plant metabolism for food security and food safety, and 4) metals uptake by organisms from their environments, including environmental contaminants. These linkages demonstrate the global need for new interdisciplinary approaches that leverage the cutting-edge capabilities of synchrotron techniques. This perspective is heavily inspired by the 2021 workshop "The Life Cycle of the Elements: rocks, soils, organisms, environment" hosted virtually at the Cornell High Energy Synchrotron Source.

## II. Synchrotron science for sustainability

### 1. Metals transport from geological processes to soils and the environment:

The metals pathway begins with planetary formation and the evolution of the deep Earth, where the oxidation states of metals and their mineral coordination and deportment provide information as to transport mechanisms through the mantle to the crust and eventually to the biosphere [16].These processes link metals distribution, remobilization and concentration to mineral resources, metals extraction, and to the areas of biosystems, organisms and human health. Key linking concepts include oxidation state characterization under diverse environmental conditions. At the high pressures and temperatures in the Earth's interior, oxygen fugacity and the oxidation states of iron and vanadium trace the role of oxygen in the deep Earth and other planetary bodies [17]. On the sea floor, oxidation states of manganese oxides reveal information of biofixation of metals and where metals sequester into particular crystal vacancy sites and determine the bio-availability of metals [18].

The cross-cutting themes of oxidation and reduction, adsorption and precipitation, and uptake mechanisms help in understanding and predicting the impacts and outcomes of the most basic interactions of metals with the environment. Because metals entering the environment originate in the Earth, understanding surface and ecosystems metal distributions reflect Earth processes that generate and recycle metals. The geological and geochemical rock-soil interface includes deep Earth to Earth's surface processes, continental crust, the seafloor, and the human ecosystem of toxins and contaminants. Thus, the diversity of mineral deposits in the crust reflects a complex history of metals source, remobilization, concentration, accessibility, and economic viability over billions of years.

Synchrotron studies of mineral resources quantitatively characterize the elemental distribution and chemical state across solid materials, both for economically valuable (e.g. gold) and deleterious (e.g. arsenic) elements [19]. Accessing metals and responsibly extracting them for future sustainable development of renewable energy infrastructure requires the thorough mineral deportment characterization of metals, a step which should be mandatory in standard mineral resources practice [20]. The importance of microscale characterization of mineral resources is reflected in the economics of mineral projects, where bulk geochemical measurement of grade (e.g. nickel) is not an accurate measure of the extractability of metals. In some cases, the metal deportment is such that the value will never be recovered and so the potential project economic returns are overestimated. In other cases, detailed understanding of the distribution of metals can help guide sustainable mining practices in sensitive environments - this goes for the seafloor as much as for the continents (Figure 1).

Once metals have been geologically concentrated in the crust and extracted for human use, the majority of materials not metallurgically concentrated end up as byproducts – as mine tailings, waste rock, or environmentally compromised soils. The element deportment of specific metals in these materials can potentially impact communities, biosystems, and industrial waste management. Although considered waste at the time of processing, today these materials are viewed as new potential sources of valuable critical minerals and metals which had not been analyzed at earlier times [21, 22]. Determination of the potentially adverse impacts on human health or the potential value in new resources in either case requires microscale characterization. Synchrotron microscale characterization can track metals in the value chain, linking metal and mineral abundances across the rock-ore-waste-soil materials flow. For example, Arduini *et al.* used synchrotron methods to document anthropogenic components in soils in an industrial area [23]. Characterization of the distribution of critical metals in soils and waste impacts future resources of the minerals as well as understanding of elements in tailings which may be economic and at the same time may be hazardous to human health.

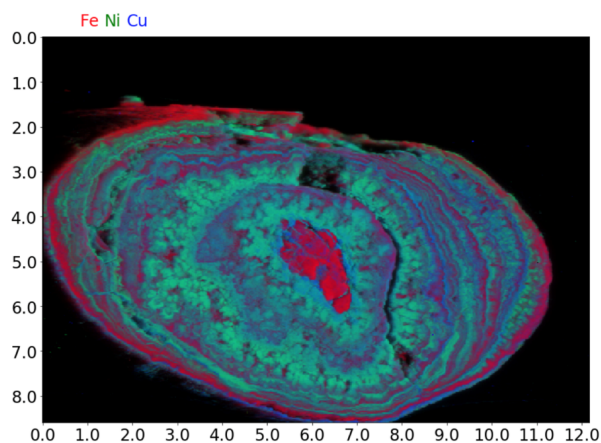

*Figure 1. Iron (red), nickel (green), and copper (blue) distributions determined by synchrotron µXRF at CHESS demonstrate materials variability within individual seafloor nodules, which may guide sustainable mining practices. Dimensions in mm. N. Mitchell, Cornell University [5].*

Metals enter waterways through weathering of both natural mineral-rock reactions [24] and anthropogenic waste [25, 26]; weathering of electronic waste is of particular environmental concern [27]. Once mobilized from rock or waste, metals-complexing in water derives from the metals' structure and adsorption characteristics, determining their reactivity and bioavailability. Together these topics link to use of microbes in mineral extraction, waste remediation and water quality [28-30].

Key questions linking metals in rocks and minerals to biogenic systems relate to redox processes. Which mineral sites are important for oxidation and reduction, and how do metals partition between these inorganic and organic sites? How do metals appear in the rock and mineral state and subsequently solubilize for transfer and microbial uptake? Questions such as these demonstrate the importance of interdisciplinary and cross-disciplinary synchrotron-based work, as key concepts link across seemingly unrelated fields such as metals from the deep Earth to human activities, biota capture and soils and water metals chemistry available for uptake.

## 2. Soil biogeochemical cycling and soil-plant interfaces

Synchrotron-based imaging and spectroscopy have been essential in establishing current understanding of elemental speciation in complex environmental systems containing interfaces between mineral surfaces, water, and biota [31, 32]. Specific examples relate to the coordination environment and bioavailability of essential micronutrients like iron and zinc, the toxicity and biotransformation of mercury, and the biogeochemical controls on arsenic solubility in the subsurface [33-35]. Now, the grand challenges for sustainability in the 21$^{st}$ Century, and the critical ecosystem services provided by soil in supporting a sustainable global agriculture system and in sequestering carbon, have re-focused attention on a new set of scientific questions probing the role of elemental cycling and speciation in these processes.

The chemical composition of soils reflects the activities of parallel and intertwined transformation processes: the weathering of parent rock material into soil minerals and the decomposition of biomass into soil organic matter. Interactions between these inorganic and organic constituents of soil – including the biogenic compounds (e.g., siderophores and other chelators) secreted into soil by microorganisms and plants – are a defining characteristic of soil and an area with open and unresolved scientific questions that underly the role of soil in sustainable agriculture and carbon sequestration. Synchrotron-based spectroscopy and tomography have enabled direct observations of spatial interactions between organic functional groups and mineral aggregates and surfaces in heterogeneous soil matrices [36], with important implications for the stabilization of organic matter in soils, e.g., the role of iron(III) oxide colloidal phases or mineral surfaces in stabilizing dissolved organic matter pools [37]. Another critical inorganic-organic interaction related to transformation of soil organic matter is the role of redox reactions of manganese and iron in the breakdown of lignocellulosic biomass. Fungi-driven mechanisms are critical in the biocatalysis of lignocellulose breakdown in the environment, utilizing enzymatic oxidation of manganese(II) to reactive manganese(III) species that can directly delignify biomass. Fungi also release organic acids that chelate iron(III) and facilitate its transport through cell walls, where subsequent redox reactions of iron lead to Fenton reactions and the production of reactive oxygen species that disrupt the lignocellulose complex [38]. While these processes have been well-characterized in simplified laboratory systems, there are open questions about the occurrence of these processes in the environment and how they are affected by redox fluctuations [39] as well as naturally occurring abundances of manganese [40] or iron [41]. Synchrotron-based techniques have been utilized to track fungal invasion of wood and the redistribution of manganese into the interior of wood by fungal hyphae [42] as well as to relate manganese speciation with decomposition of plant litter [43]. These interactions between inorganic and organic constituents of complex soil matrices ultimately govern the mineralization of organic matter to carbon dioxide [44], and there is growing evidence that rates of breakdown of lignin-rich biomass into labile carbon can also regulate nitrogen metabolism at terrestrial-aquatic interfaces [45].

The importance of inorganic-organic interactions extends to understanding of the bioavailability and toxicity of trace elements in soils, via complexation reactions between trace elements and (biogenic) organic ligands as well as redox-active properties of biogenic low molecular weight organic molecules secreted by bacteria. Coordination reactions govern the solubility and plant-availability of essential (micro)nutrients, as in the well-known case of siderophores [46, 47], but binding of trace elements with ligands in natural organic matter can also control the bioavailability (and toxicity) of toxins. For example, molecular-level synchrotron studies have shown that sulfhydryl functional groups in natural organic matter can sequester arsenite [48, 49], mercury [50], cadmium [51], and other toxic trace elements. There is notable recent evidence that redox-active biogenic compounds secreted by bacteria can liberate orthophosphate from mineral surfaces by accelerating the reductive dissolution of iron oxide mineral phases [52]. The importance of these processes in heterogeneous soil matrices underscore the need to couple µXRF imaging with spectroscopy using both hard and soft X-rays. Carbon NEXAFS (280-310 eV) with STXM have provided information on spatial distribution of organic functionalities associated with mineral aggregates [53]. Because of the high affinity of "soft" metals and metalloids (e.g., cadmium; mercury; arsenic) for thiols, the combination of sulfur XANES with spectroscopy of heavier elements has provided new insights into the role of organic sulfur speciation in controlling the mobility of critical trace elements [48, 54].

Soils are highly dynamic and living systems. Perturbations to soil physical-chemical properties occur on the time scale of minutes to hours, as in the case of soil flooding and consequent changes to soil redox conditions, to the scale of weeks to months, as in the case of plant and plant-mediated elemental uptake and translocation. Prior imaging of rice rhizospheres and iron plaque formation on rice roots represent a discrete snapshot in time [10, 55]. While new approaches have been developed for mapping spatio-temporal variations in rhizosphere processes [12], the ability to track these changes in real-time using synchrotron-based techniques would support a new level of detail in understanding these systems (Figure 2).

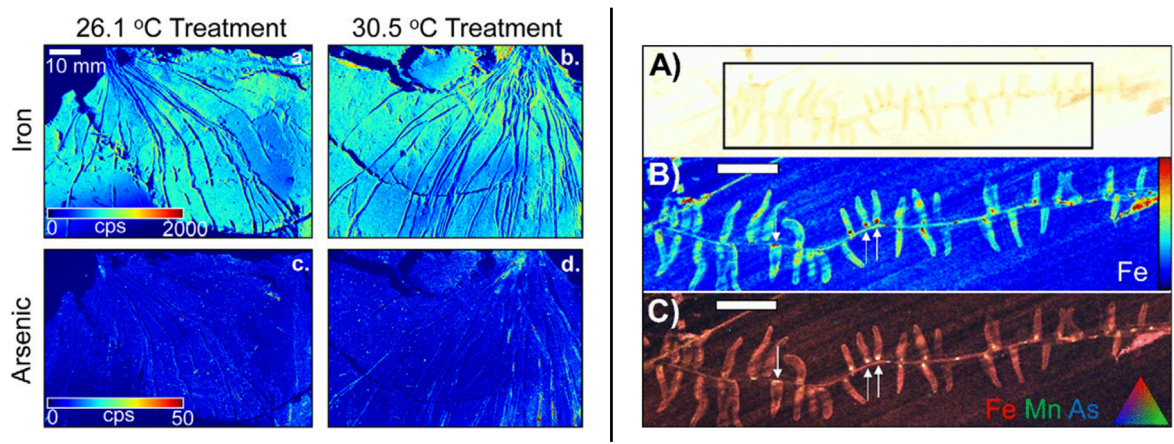

*Figure 2. Plant-root-soil interactions probed by synchrotron µXRF. Left: µXRF imaging of rice roots grown in rhizobox soils, demonstrating impact of increased temperatures on arsenic distribution; reprinted with permission from [10], copyright 2017 John C. Wiley and Sons. Right: µXRF imaging of iron-rich plaque residues from rice roots, showing microscale localization of arsenic; scale bar is 5 mm; reprinted with permission from [12], copyright 2021 American Chemical Society. The ability to track such systems in real time using synchrotron methods would represent a major advance.*

## 3. Metals in plant metabolism for food security and food safety

Metals such as zinc, iron, copper, cobalt, nickel, manganese, magnesium and calcium play critical structural and chemical roles as cofactors in a wide range of essential biological processes, including respiration, scavenging reactive oxygen species, transcription, photosynthesis, signal transduction and hormone perception. Owing to these and other essential cellular functions, it is not surprising that more than 25% of proteins are thought to need metals [56, 57]. Considering the essential yet toxic nature of transition metals, a balance between different metal homeostasis factors including metal-specific importers and exporters, metal storage proteins, ligands, metallochaperones, and metal-sensing transcriptional regulators, is required to ensure sufficient metal supply for the correct cellular function while avoiding metal overload [58, 59]. However, much of the cellular infrastructure that ensures proper metal homeostasis remains poorly characterized.

Considering that metal imbalance leads to various disease states in humans and limits plant/crop development that in a field setting translates to reduced yields, poor nutritional quality and mineral nutrient density in edible plant tissues, the identification of factors contributing to metal homeostasis is important for the future of global food security and human health. Concerning plants, their ability to acquire and accumulate mineral nutrients in edible tissues can be severely compromised by extreme weather conditions that are increasing in frequency and intensity, further hampering biofortification efforts.

Functional genomic tools utilized over the last decade and the emergent CRISPR-Cas-mediated genome editing approaches facilitated the identification and characterization of metal homeostasis relevant genes. These tools provide an excellent entry point for understanding the complexity and hierarchy of metal homeostasis networks. The picture is not complete, however, without the ability to visualize metals, quantify and probe them, in the surrounding chemical environment (oxidation state, local coordination geometry) at molecular, subcellular, cellular, tissue, and organ levels. Total metal concentrations averaging over an entire tissue type may not distinguish metal distributions within tissues [60]. When feasible, 2D-XRF in confocal mode (2D-CXRF) with a specialized x-ray collection optic is preferable to traditional XRF methods (both 2D-XRF and 3D micro-XRF tomography) for measuring spatially-resolved total metal concentrations because it allows quantitative volumetric comparisons of metal distributions (as opposed to concentration per area as in 2D XRF) among different samples, without the need to limit sample thickness or lateral size (as in XRF tomography) [61]. Using 2D-CXRF enabled a micron-scale resolution capture of differences in iron, copper, zinc and manganese localization and concentrations in different plant cell types because of genetic perturbations [7, 62]. Figure 3 shows the complementarity of conventional whole-tissue 2D XRF and quantitative 2D-CXRF.

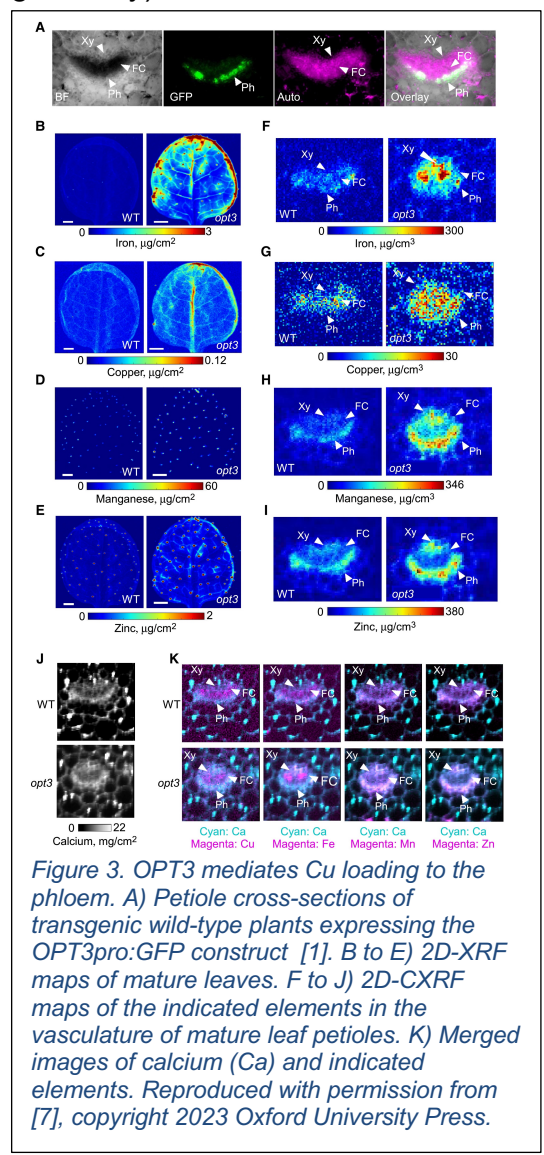

*Figure 3. OPT3 mediates Cu loading to the phloem. A) Petiole cross-sections of transgenic wild-type plants expressing the OPT3pro:GFP construct [1]. B to E) 2D-XRF maps of mature leaves. F to J) 2D-CXRF maps of the indicated elements in the vasculature of mature leaf petioles. K) Merged images of calcium (Ca) and indicated elements. Reproduced with permission from [7], copyright 2023 Oxford University Press.*

It is also critical to be able to link local metal concentration and oxidation state with specific cell types or local expression of a specific gene. X-ray absorption near-edge structure (XANES) imaging bridges the gap between imaging of the spatial distribution of elements and their speciation [15]. Linking these two synchrotron techniques in multimodal operation is especially important for understanding the biology and metabolic functions resulting from the distribution of elements that easily undergo the oxidation-reduction reactions (e.g. iron, copper, manganese, arsenic). Combining multimodal synchrotron-based metal imaging with the optical (brightfield/fluorescence) microscopy for the simultaneous study of protein localization and morphological features can serve an ultimate role in bridging the genotype-to-phenotype gap and assign function in mineral homeostasis to unknown plant proteins (Figure 3A).

In addition, it is increasingly important to understand how plant homeostasis mechanisms respond to external stimuli, from pathogens to environmental stresses. Time-resolved optical imaging of growing plants serves as a form of dynamic phenotyping which can clarify homeostasis mechanisms by comparing wild type and knockout plants [8, 9] (Figure 4). The ability to image temporal dynamics of metals in living plants under different genetic and environmental perturbations would be a major step forward for this research area. There is a need for dedicated synchrotron-based infrastructure to take on these challenging measurements, including supporting live plants, providing environmental controls for perturbations, all while minimizing the x-ray dose to the plant to enable time series measurements.

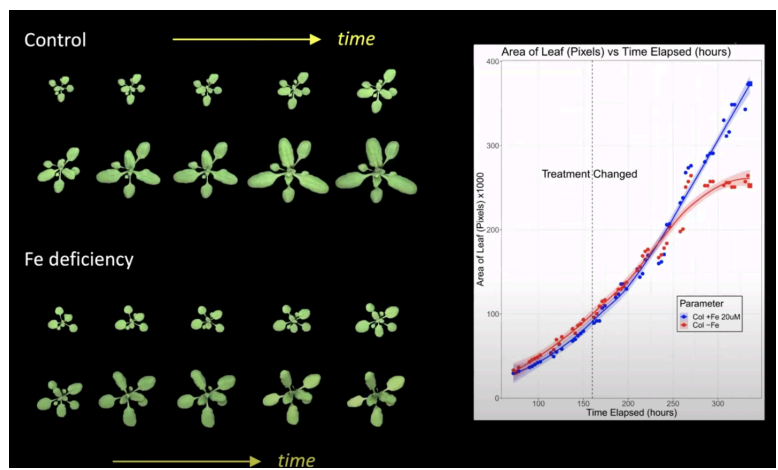

*Figure 4. Dynamic optical phenotyping during Fe deficiency captures previously undescribed transient phenotypes. There is a need for an elemental counterpart to this type of exeriment, i.e. dynamic elemental phenotyping during perturbations to plant environment [8, 9].*

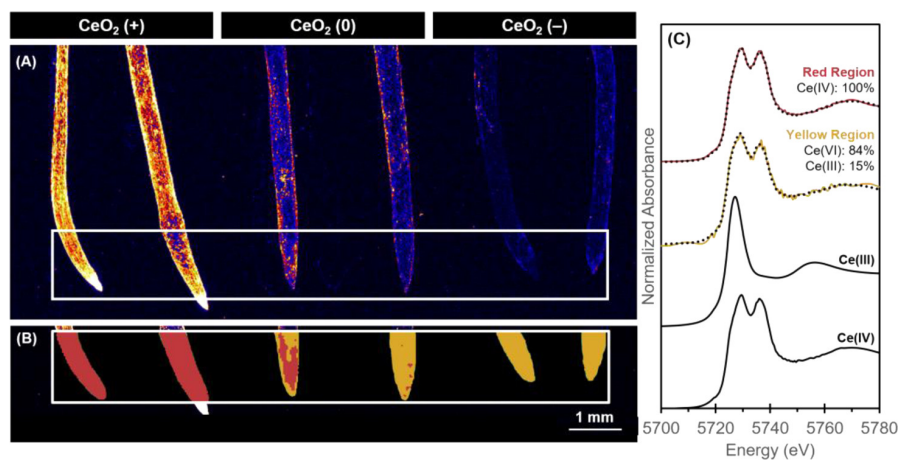

*Figure 5. Nanoparticle interactions with plant roots: Wheat roots exposed hydroponically to 20 mg cerium/L positive (+), neutral (0), or negative (−) coated $CeO_2$ nanoparticles for 34 hours. (A) μXRF map showing total cerium distribution, brighter colors indicate higher Ce concentrations. (B) The spatial distribution of two pixel populations identified (red and yellow) from XANES mapping (white box in A). (C) Normalized Ce LIII XANES spectra corresponding to the two pixel populations (red and yellow) plus the spectra for the reference compounds. The black dotted lines are fitted data while the solid lines are experimental data. Reproduced with permission from [4], copyright 2017 American Chemical Society. Improved infrastructure for plant growth as well as access to real-time dynamic or discontinuous periodic measurements at synchrotrons would enable future research.*

Furthermore, an emerging topic in the context of agricultural sustainability that engages directly with elemental speciation is the use of engineered nanomaterials to deliver micronutrients or other substances to plants. Existing applications of agrochemicals such as fertilizer to crop soils or leaves are estimated to be highly inefficient, with the majority of applied materials not taken up by plants, leading to tremendous waste and environmental impacts, e.g. through runoff [63]. Nanomaterials hold the promise of purposeful engineering of plant-nanoparticle interactions, potentially delivering what a plant needs when and where it is needed [64]. Many questions surround plant-nanoparticle interaction, including mechanisms of uptake by roots and leaves [4, 65, 66], impacts on crop nutrition [67-69], as well as clarification of conditions where nanoparticles function more efficiently than conventional agrochemicals [70, 71]. Synchrotron methods are essential for spatial analysis of elemental distributions and chemical state changes [72], and time dependent XRF and XAS studies on the microscale are needed to track the movements and transformations of nanoparticles through soils and within living plants (Figure 5).

## 4 Metals in animal metabolism and insights into ecosystem health

X-ray methods for imaging and spectroscopy have established new frontiers of study in the ways animals take up and use metals in their environments. Imaging of sequentially deposited tissues can provide detailed lifetime exposure history as well as insight into environmental trends. For instance, keratin-rich tissues such as goat horn can record exposures to lead [73], and impacts on human diet of arsenic content and chemical form in dried baby shrimp can be evaluated [74]. Measuring and mitigating the anthropogenic impacts on animals and ecosystems will require new understanding of the factors influencing metal bioavailability in specific ecosystems.

Ocean acidification and hypoxia threaten keystone marine invertebrates, destabilizing fisheries and coastal food webs around the world. Fish play an essential role in global nutrition strategies, providing animal protein for billions of people worldwide with less environmental impact than land-based livestock. However, marine fish are a significant dietary source of mercury worldwide. There is an urgent need to understand the fundamental mechanisms of mercury uptake by fish and other aquatic organisms, as well as the mechanisms of mercury-selenium interactions, both to mitigate ecotoxicity and to safeguard food sources [2, 75, 76]. Selenium uptake alone can also have profound effects on aquatic life whose mechanisms are not fully understood [77, 78]. The complex relationships between metal bioavailability, toxicity, and surrounding water chemistry, especially in brackish and seawater, demand further study. Future advances in this area will require both high-spatial-resolution elemental mapping and high-energy-resolution x-ray spectroscopy (e.g. high energy resolution fluorescence detected XAS, or HERFD–XAS) to understand distribution and speciation on an intracellular scale (Figure 6).

On a broader scale, how can we monitor and mitigate changes to ecosystem health? The distribution of the trace element manganese in fish ear stones (otoliths) has been shown to record exposure to low-oxygen aquatic environments [79-82]. Detailed elemental mapping of these hard tissues provides markers for environmental exposures over the lifetimes of individuals, yielding new insights into ocean environmental changes, migratory patterns, and impacts of human developments on environments [83, 84]. To achieve a more nuanced understanding of the environmental records in individual organisms moving forward, it will be important to understand the mechanisms for elemental deposition in hard animal tissues, including other types such as corals, shells, and horn.

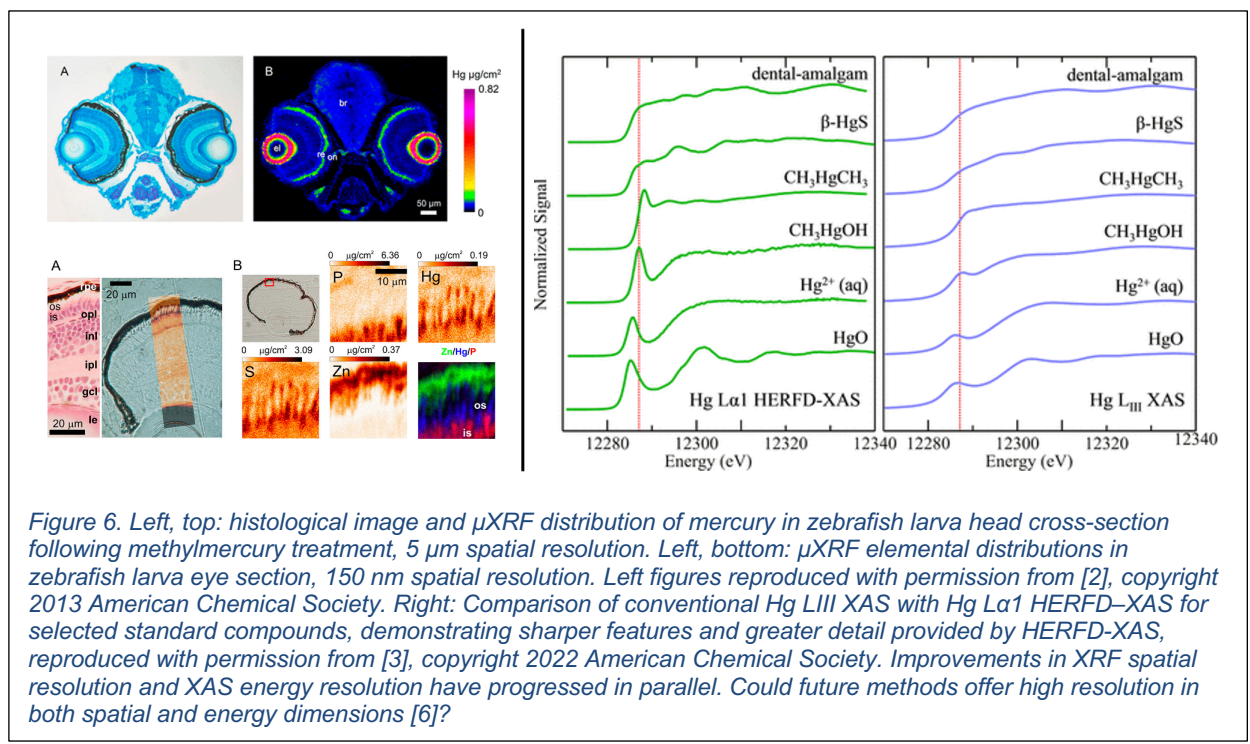

*Figure 6. Left, top: histological image and µXRF distribution of mercury in zebrafish larva head cross-section following methylmercury treatment, 5 µm spatial resolution. Left, bottom: µXRF elemental distributions in zebrafish larva eye section, 150 nm spatial resolution. Left figures reproduced with permission from [2], copyright 2013 American Chemical Society. Right: Comparison of conventional Hg LIII XAS with Hg Lα1 HERFD–XAS for selected standard compounds, demonstrating sharper features and greater detail provided by HERFD-XAS, reproduced with permission from [3], copyright 2022 American Chemical Society. Improvements in XRF spatial resolution and XAS energy resolution have progressed in parallel. Could future methods offer high resolution in both spatial and energy dimensions [6]?*

Human impacts on the environment are closely linked to other areas in the life cycle of metals. Nanoparticle-based agricultural treatments promise improved crop resilience, but environmental impacts must be evaluated, as the same treatments may disrupt metal homeostasis in terrestrial [85] or marine organisms [86]. Furthermore, an enormous fraction of anthropogenic mass is plastic, which is known to persist in the environment as micro- and nano-plastic particles. Many questions surround these micro- and nano-plastics; for instance, how do plastic particles act as vectors for inorganic contaminants, in water, soils, plants, and animals, and what is the bioavailability- in different organisms- of the metals that are present [87-89]?

## III. Experimental needs, design, and outlook

X-ray tools are ideally suited to answer numerous questions relating to the life cycle of metals, where a common theme is the need to inventory elemental and chemical composition in both space and time. Two crucial tools for such inventories can be broadly categorized as synchrotron x-ray fluorescence (XRF) and x-ray absorption spectroscopy (XAS), but complementary methods using both x-rays (e.g. x-ray diffraction, scanning transmission X-ray microscopy) and additional optical probes (e.g. infrared and Raman spectroscopy, reflectance spectroscopy, fluorescence microscopy, and mass spectrometry techniques) are also valuable [14]. An ideal facility to support life cycle of metals questions would provide a versatile suite of x-ray methods and complementary optical spectroscopy tools for characterizing static samples (ex-situ) as well as advanced capabilities for in-situ, time-resolved measurements.

1. Ex-situ measurements

2D XRF imaging with focused x-ray beams is a powerful tool for measuring quantitative elemental distributions at micro- and nano-scale resolution across a wide range of length scales and a broad range of sample types [31, 32, 72, 90-92]. It is often desirable to measure elemental distributions across a macroscopic or organism-scale area, and then select a region of interest for more detailed examination, at higher resolution and/or with additional probes. There is a need for facilities that can provide elemental imaging at a broad range of spatial resolutions (>100 μm to <100 nm, i.e. whole organism to subcellular) and scan areas (>10 cm to <1 mm) for the same sample. Large-area surveys need to be rapid, both to minimize potential for sample damage and to efficiently define the experimental path forward. At the same time, there is a need for high elemental sensitivity, as many questions are related to low and ultra-low concentrations (few ppm and below). For certain high atomic number elements (e.g. rare earth elements, cadmium, iodine), incident x-ray energies above 25 keV are desirable to excite K-line fluorescence, avoiding signal overlaps and providing higher sensitivity, but most micro-focusing x-ray beamlines have a high energy cutoff around 25-30 keV. There is also a need for improvements to cyberinfrastructure, including data visualization tools to match the acquisition speed, including real-time spectrum fitting, and user-friendly image analysis tools to find elemental correlations and detect anomalies.

Coupled with elemental inventory is the need for details related to chemical composition. XAS, typically a bulk or single point measurement, can identify an element's oxidation state and provide insights into its bonding environment [72, 92, 93]. Since XAS measurements typically involve continuous x-ray exposure over seconds, facilities must work to prevent sample damage through sample environments (e.g. cryogenic temperatures and inert atmospheres) as well as instrumentation to reduce the total exposure time (e.g. fast monochromator scanning, increased efficiency of fluorescence detection). As with 2D XRF, facilities are needed to enable XAS data collection at multiple length scales to understand sample heterogeneity. XAS mapping can support this aim, either by collecting XAS spectra point by point, or by sequentially scanning the same area at different incident x-ray energies [93].

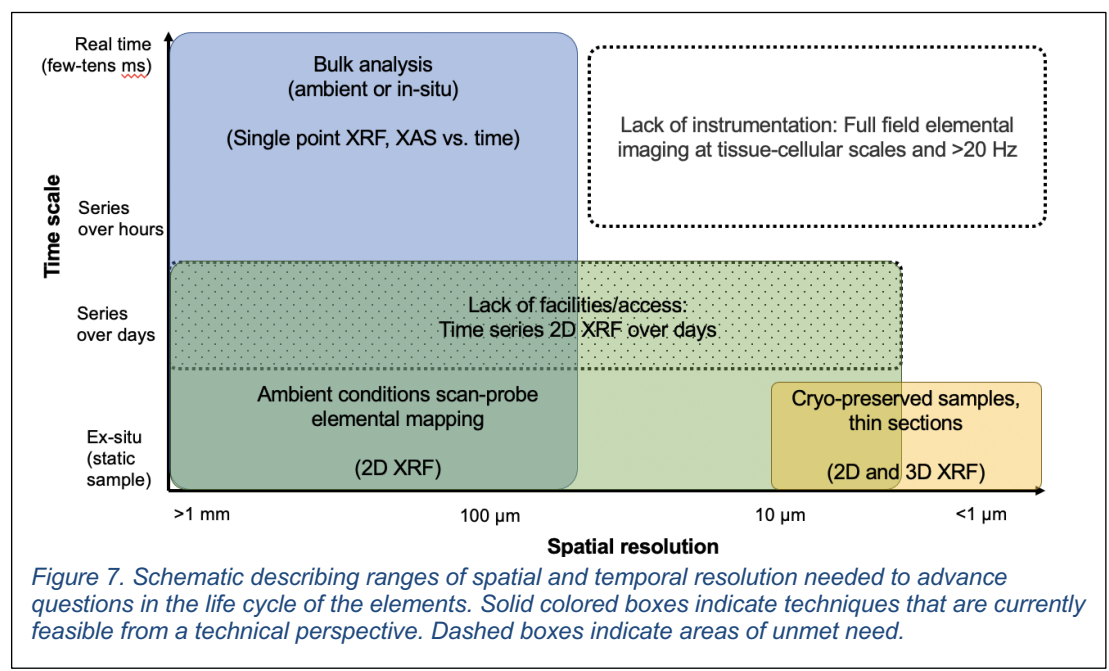

*Figure 7. Schematic describing ranges of spatial and temporal resolution needed to advance questions in the life cycle of the elements. Solid colored boxes indicate techniques that are currently feasible from a technical perspective. Dashed boxes indicate areas of unmet need.*

3D XRF is another core tool for ex-situ elemental inventory. XRF tomography is an effective tool for 3D microscale visualization of element distributions [60, 94, 95], but the spatial resolution of the reconstructed images is limited by sample thickness and the total x-ray exposure time is high, making this method most appropriate for small, dehydrated samples such as seeds. Confocal XRF is a valuable approach for measuring 3D elemental distributions in samples whose shape is not suited to tomography, as well as for measuring depth-specific XAS [96, 97].

2. In-situ measurements

Many questions related to the movement of elements through the environment and organisms will require time-resolved work on multiple spatial and temporal scales. The time scale of interest tends to shrink with decreasing length scale, as one might intuit from the diffusion constant units of area/time. This parameter space is schematically shown in Figure 7 and discussed below.

On the longer time scale, the ex-situ capabilities described above could be leveraged to perform periodic measurements of slow processes (days to weeks). For example, monitoring chronic exposure experiments or real-time growth processes such as seed development or germination could be enabled by making periodic measurements on the same sample or system over time. Such experiments do not fit into typical beamtime slots, and would require innovations to facility access models. On a more moderate time scale (minutes to hours), time-resolved XAS could be applied to in-situ systems such as *operando* flow cells for microscale adsorption measurements in soils. Efforts towards in-situ XRF measurements of live plants, enabled by high count rate detectors such as the Maia, have begun to establish radiation dose thresholds for hydrated plant tissue [98, 99]. Dynamic XRF imaging of iodine and gold nanoparticles in live plants has also been demonstrated, taking advantage of reduced radiation doses using unfocused, high-energy (53 keV) x-rays [100]. However, these experiments remain challenging to implement routinely, not only due to the technical challenges of minimizing dose while maximizing signal, but also due to a lack of infrastructure to support live plants under environmentally controlled conditions at synchrotron facilities.

On fast time scales (ms to seconds), it would be extremely valuable to observe elemental movements in the vasculature of an organism in real time, such as elemental loading and unloading in phloem [8, 65], or flow of hemolymph through insect wings [17]. Even smaller and faster imaging would enable access to elemental movements within cells. These experiments could potentially one day be enabled by full field XRF imaging with emerging "color x-ray camera" detectors [101-104].

3. Complementary x-ray, optical spectroscopy, and mass spectrometry probes

Especially when examining geological and soil specimens, x-ray diffraction (XRD) is a powerful complementary tool to XRF and XAS, as it can provide unambiguous mineral phase identification [105]. It is possible to perform XRF and XRD imaging simultaneously, especially on thin samples where transmission XRD is convenient [106, 107].

Non-x-ray optical probes, such as infrared, UV-Visible, or Raman spectroscopy, can also provide valuable complementary information to XRF and XAS [108, 109]. Such spectroscopy can aid in identification of light-element/organic compounds as well as in phase identification of dense materials when transmission XRD is not practical. It would be valuable for a microprobe x-ray facility to incorporate one or more of these probes into its suite of routine tools, in parallel or simultaneous with x-ray imaging tools. At tissue- and cell-specific length scales, it would be useful to localize proteins & elements simultaneously, which could be accomplished with fluorescence microscopy of green fluorescent protein tags [110], for example.

Mass spectrometry techniques such as laser-ablation coupled with inductively coupled mass spectrometry (LA-ICP-MS) and high-resolution secondary ion mass spectrometry (SIMS, nanoSIMS) are highly complementary methods to synchrotron XRF imaging for studying geological [111] and biological specimens [112-114]. In particular, nanoSIMS provides access to nanoscale distributions of low-atomic number element isotopes such as carbon, nitrogen, oxygen, phosphorus, and sulfur; while synchrotron XRF excels at measuring micro- to nano-scale distributions of higher atomic number trace elements. Researchers would benefit from access to mass spectrometry imaging tools in conjunction with synchrotron imaging methods.

4. Sample Handling

Sample preparation is a critical aspect of a successful synchrotron measurement, with the aim of studying a chemical state which accurately represents the "real" context of interest. A strength of many of the µXRF and other x-ray techniques described in this perspective is that minimal sample preparation is required and measurements may be taken in ambient temperatures and atmospheres. However, for hydrated tissues and other sensitive samples, high-intensity focused x-ray beams can induce sample damage. These issues are most severe for experiments requiring highly focused beams (e.g. nano-XRF) and/or long exposures in a single spot (e.g. XAS, 3D XRF tomography). It is therefore sometimes important to be able to work with frozen-hydrated, possibly cryo-sectioned samples; this of course requires beamline infrastructure to support scanning of frozen samples, and ideally sample preparation facilities. On the other hand, as high-count-rate detectors continually improve signal collection efficiency, the total dose may be low enough to avoid perturbing hydrated or even living tissues, a regime which would be relevant for the dynamic studies discussed in this perspective. In this case, an ideal facility would provide access to efficient, low-per-pixel dwell time µXRF scanning, as well as resources to enable plant and/or animal growth both on and off the beamline.

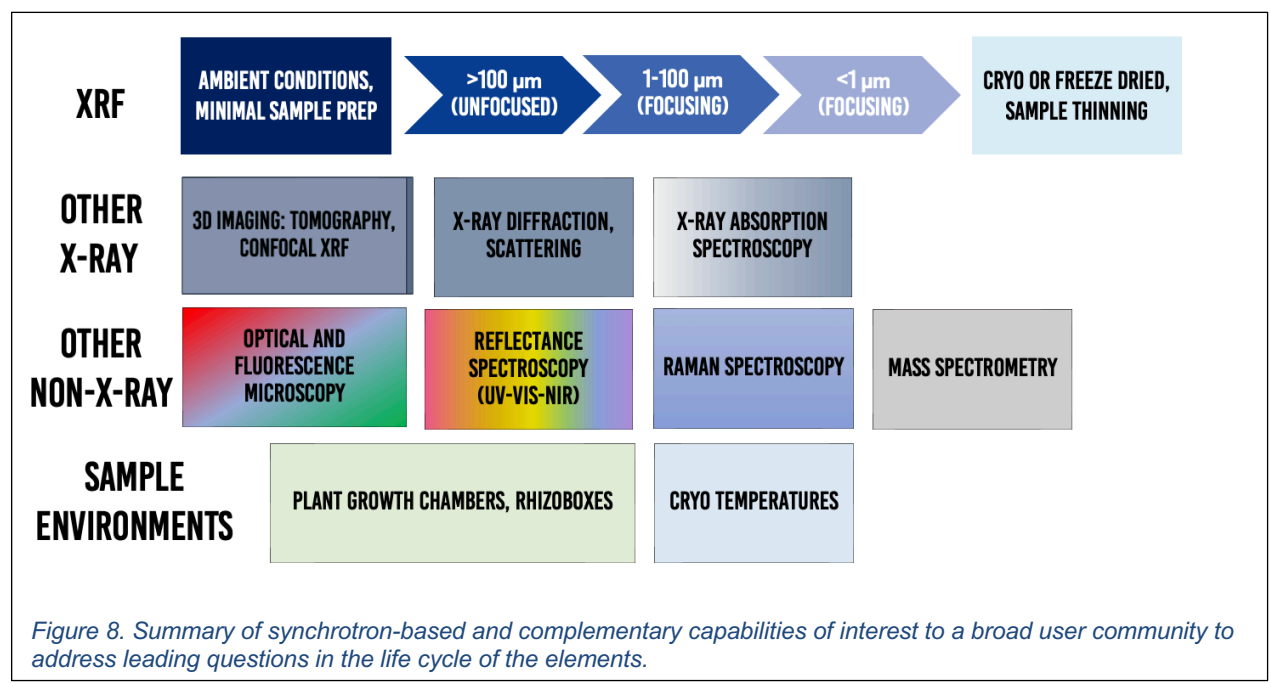

*Figure 8. Summary of synchrotron-based and complementary capabilities of interest to a broad user community to address leading questions in the life cycle of the elements.*

## Conclusions

Understanding the mechanisms of elemental movement across inorganic and organic, geological and biological spheres is critical to developing new strategies for human sustainability in the face of dwindling resources and a changing global climate. Urgent responses are needed to measure, quantify, and address solutions to our collective challenges, including the sustainable extraction and reuse of critical metals; global food safety, security, and nutritional quality; and minimizing human impacts on the environment.

Careful attention to quantifying metals deportment, chemical state, bioavailability, mobility, uptake, and mechanisms of cycling and homeostasis, is therefore necessary across a wide range of scientific fields. From the rocks and soil beneath our feet, to the soil we cultivate, and onwards to the plants, animals, and water we consume, we are surrounded by pressing questions about the nature of elemental pathways and transformations. The common thread of the life cycle of metals links these seemingly disparate fields in a common quest to document, describe, understand, and derive knowledge from observations of the natural world in order to address the critical sustainability issues of our times. We consider this approach to be a high-priority need for researchers to address global issues.

Synchrotron-based methods have the potential to enable the next generation of researchers to ask and answer research questions with increasing richness of both fundamental understanding and societal impact. We have identified some specific technical needs which would help advance questions in the life cycle of metals. First, there is enormous demand for a highly versatile, multi-scale, multi-probe, micro-focusing synchrotron beamline covering 2D and 3D µXRF as well as time- and spatially resolved µXAS and µXRD. An ideal facility would also offer complementary imaging modalities such as infrared, Raman, or reflectance spectroscopy, as well as access to mass spectrometry- in other words, a "one stop shop" for metals inventory in a wide range of geologically, environmentally, and biologically relevant samples (Figure 8). Second, there is a strong emphasis from the community on moving towards dynamic experiments as opposed to exclusively static samples, with a vast range of length and time scales of interest. Time-resolved 2D µXRF measurements will require, on the longer hours-days scale, management and mitigation of radiation dose, as well as creative access models to enable non-sequential measurements; and on the aspirational shorter time- and length-scales, new developments in detector technology and imaging approaches are needed.

Finally, there is enormous untapped potential in synchrotron studies of the life cycle of metals, even with the present state of technical capabilities. In addition to their quantitative meaning, synchrotron chemical imaging methods provide a powerful tool for science communication, and can help the public understand the source, cause and effects of environmental impacts. Further, quantitative studies of metals in natural systems could be better linked across fields to improve the holistic picture of how metals move and transform as they pass through inorganic and organic systems. There is a need for training and workshops to enable more researchers to take advantage of synchrotron tools and communicate across traditional field boundaries.

### Acknowledgements

This perspective piece was motivated in part by the lively discussion of presenters and participants at the July 2021 workshop, "The life cycle of the elements: Rocks, soils, organisms, environment (X-LEAP)- a CHESS 2030 workshop." This Zoom workshop attracted 123 registered participants and was supported by the National Science foundation (MCB-2132738). The workshop was hosted by the Cornell High Energy Synchrotron Source (CHESS) and the Center for High-Energy X-ray Sciences at CHESS (CHEXS), which is supported by the National Science Foundation under award DMR-1829070.

### Data Availability

Raw data for images in this article were generated at multiple light sources including CHESS (Figures 1, 3), the Stanford Synchrotron Radiation Lightsource (SSRL, Figures 2, 6), the National Synchrotron Light Source II (NSLS-II, Figures 2, 3), the Australian Synchrotron (Figure 5), and the Advanced Photon Source (Figure 6). Derived data may be available from the corresponding authors referenced in each figure on request.